\documentclass[12pt]{article}

\usepackage{graphicx}

\def\um{{1}/{2}}
\def\sq{1/\sqrt{2}}

\begin{document}
\renewcommand{\thefootnote}{\fnsymbol{footnote}}
\renewcommand{\theenumi}{(\roman{enumi})}
\title{Neutrinoless double $\beta$ decay, 
neutrino mass hierarchy, and neutrino dark matter}
\author{
{Naoyuki Haba$^{1,2}$}\thanks{haba@eken.phys.nagoya-u.ac.jp}
{, Noboru Nakamura$^2$}\thanks{nakamura@eken.phys.nagoya-u.ac.jp}
{, and Tomoharu Suzuki$^2$}\thanks{tomoharu@eken.phys.nagoya-u.ac.jp}
\\
\\
\\
{\small \it $^1$Faculty of Engineering, Mie University,}
{\small \it Tsu Mie 514-8507, Japan}\\
{\small \it $^2$Department of Physics, Nagoya University,}
{\small \it Nagoya, 464-8602, Japan}\\
}
\date{}
\maketitle

\vspace{-9.5cm}
\begin{flushright}
hep-ph/0205141\\
DPNU-02-12\\
\end{flushright}
\vspace{9.5cm}
\vspace{-2.5cm}
\begin{center}
\end{center}
\renewcommand{\thefootnote}{\fnsymbol{footnote}}

\begin{abstract}

Recently 
 the evidence of the neutrinoless double $\beta$ 
 ($0\nu \beta\beta$) decay has been announced. 
This means that neutrinos are Majorana particles and 
 their mass hierarchy is forced 
 to $m_1\sim m_2\gg m_3$ (Type B) or 
degenerate mass, $m_1\sim m_2\sim m_3$ (Type C) 
 patterns in the diagonal basis of
 charged lepton mass matrix, where $m_i$ is the absolute value of ${\it
 i}$-th generation neutrino mass. 
We analyze the magnitude of $0\nu \beta\beta$ decay 
 in the degenerate neutrinos including 
 the cosmological constraint of 
 neutrino dark matter, since 
 the Type B mass hierarchy pattern
 always satisfies the cosmological 
 constraint.  
The upper bound of neutrino absolute mass 
 is constrained by $0\nu \beta\beta$ experiment or 
 cosmology depending on Majorana $CP$ phases
 of neutrinos and solar mixing angle.

\end{abstract}

\newpage

Recently the evidences of neutrino oscillations are strongly supported
 by both of the atmospheric~\cite{Kamiokande,SKatm} 
 and the solar neutrino experiments~
\cite{SKsolar,SNO,Cl-Homestake,Ga-Gallex-and-GNO}. 
The former suggests an almost maximal lepton flavor mixing 
 between the 2nd and the 3rd generations, 
 while the favorable solution to the solar neutrino deficits 
 is given by large mixing angle solution between the
 1st and the 2nd generations ( LMA, LOW  )~\cite{SKsolar}\cite{post-SNO-analysis}\cite{SNO1}.
Neutrino oscillation experiments indicate that the neutrinos have tiny
 but finite masses, with two mass squared differences 
 $\Delta m_{\odot}^2< \Delta m_{\rm atm}^2$.  
However, 
 we cannot know 
 the absolute values of the neutrinos masses 
 from the oscillation experiments. 

Recently, 
 a paper\cite{db} 
 announces the evidence of 
 the neutrinoless double $\beta$ 
 ($0\nu \beta\beta$) decay.  
This paper suggests 
\begin{eqnarray}
\langle m \rangle=(0.05-0.86) \mathrm{eV}\quad at\;97\%\;\mathrm{c.l.}
\quad\mathrm{(best\;value\;0.4\;eV)}\;.
\label{result}
\end{eqnarray}
This result is very exciting\footnote{
There are arguments for this result in Refs.\cite{bb}.
}.
It is because $0\nu\beta\beta$ decay experiments
 could tell us about the absolute value of the neutrino masses, 
 while neutrino oscillation experiments 
 show only mass squared differences of neutrinos. 
Some papers \cite{OTHERS} have studied from this date. 
The evidence for $0\nu\beta\beta$ decay 
 also means neutrinos are the Majorana particles and 
 the lepton number is violated, since 
 $0\nu\beta\beta$ decay cannot occur in the case of 
 Dirac neutrinos. 
This evidence is also closely related to 
 the recent topics of the cosmology such as 
 the dark matter candidate of universe\cite{dark}. 

The tiny neutrino masses and 
 the lepton flavor mixing have 
 been discussed in 
 a lot of models beyond the Standard Model (SM). 
One of the most promising ideas is 
 that light neutrinos are constructed as Majorana particles 
 in the low energy, such as 
 the see-saw mechanism\cite{seesaw}. 
Here we are concentrating on 
 the light Majorana neutrinos 
 which masses are induced by the 
 dimension five operators in the 
 low energy effective Yukawa interactions. 

In the previous paper\cite{HS}, 
 we have estimated the magnitude of $0\nu \beta\beta$ decay
 in the classification of the neutrino mass hierarchy patterns 
 as Type A, $m_{1,2} \ll m_{3}$, Type B, 
 $m_1 \sim m_2 \gg m_3$, 
 and Type C, $m_1 \sim m_2 \sim m_3$,
 where $m_{i}$ is the absolute values of 
 the $i$-th generation neutrino \cite{Altarelli}. 
The magnitude of $0\nu \beta\beta$ decay 
 strongly depends on the neutrino mass hierarchy. 
According to the analysis in Ref.\cite{HS}, 
 the results of $0\nu \beta\beta$ suggest 
 that neutrino mass hierarchy is forced 
 to Type B or Type C
 patterns in the diagonal basis of
 charged lepton mass matrix.

In this paper we will 
 estimate the magnitude of $0\nu \beta\beta$ decay 
 in the degenerate neutrinos including 
 the cosmological constraint of 
 neutrino dark matter. 
The relation of $0\nu \beta\beta$ decay and 
 neutrino dark matter has been analyzed 
 in Ref.\cite{dark}. 
We will analyze this relation 
 including mass squared differences 
 of the solar LMA solution and the value of $U_{e3}$ 
 in accordance with Majorana phases of neutrinos 
 masses. 
Since the Type B mass hierarchy pattern
 always satisfies the cosmological 
 constraint, $\Sigma = m_1 + m_2 + m_3 \leq 4.4$ eV\cite{cosmo}, 
 we will be concentrating on degenerate neutrino masses, 
 Type C.  
We will see 
 the upper bound of neutrino absolute mass 
 is constrained by $0\nu \beta\beta$ experiment or 
 cosmology depending on Majorana $CP$ phases
 of neutrinos and solar mixing angle.
\\   
%
%
%
%
%

In the diagonal base of the charged lepton sector, 
 the light neutrino mass matrix, $(M_\nu)_{ij}$, 
 is diagonalized by $U_{ij}$
 as
\begin{eqnarray}
U^{\mathrm{T}}M_\nu U \equiv M_\nu^{diag}
=diag(m_1,m_2,m_3)\quad.
\label{MM}
\end{eqnarray}
Here the matrix $U$ is so-called MNS matrix\cite{MNS} 
 denoted by 
\begin{eqnarray}
U=V \cdot P\quad,
\label{MNSP}
\end{eqnarray}
where $V$ is the CKM-\textit{like} matrix,
which contains one $CP$-phase ($\delta$),
\begin{eqnarray*}
V=\left(
\begin{array}{ccc}
c_{13}c_{12}&c_{13}s_{12}&s_{13} e^{-i\delta}\\
-c_{23}s_{12}-s_{23}s_{13}c_{12}e^{i\delta}&
c_{23}c_{12}-s_{23}s_{13}s_{12}e^{i\delta}&
s_{23}c_{13}\\
s_{23}s_{12}-c_{23}s_{13}c_{12}e^{i\delta}&
-s_{23}c_{12}-c_{23}s_{13}s_{12}e^{i\delta}&
c_{23}c_{13}
\end{array}
\right)
\end{eqnarray*}
and 
$P$ contains two extra Majorana phases ($\phi_{1,2}$),
\begin{eqnarray*}
P=diag.(e^{-i {\phi_1/2}},e^{-i \phi_2/2},1)\quad.
\end{eqnarray*}
For the suitable classification, 
 we introduce the matrix 
\begin{eqnarray}
\widetilde{M}_\nu\equiv 
P^* M_\nu^{diag} P^{*}=diag.(m_1e^{i \phi_1},m_2 e^{i\phi_2},m_3)
=diag.(\tilde{m}_1,\tilde{m}_2,\tilde{m}_3)\quad.
\label{tildeM}
\end{eqnarray}

The results from 
 the recent neutrino oscillation experiments\cite{
Kamiokande,SKatm,SKsolar,SNO,Cl-Homestake,
Ga-Gallex-and-GNO,post-SNO-analysis}  
 indicate that the neutrinos have tiny
 but finite masses, with two mass squared differences 
 $\Delta m_{\odot}^2< \Delta m_{ atm}^2$.  
The naive explanation of 
 the present 
 neutrino oscillation 
 experiments is that 
 the solar neutrino anomaly is caused by 
 the mixing of 
 the 1st and the 2nd generations 
 ($\theta_{\odot}\simeq\theta_{12}$, 
 $\Delta m^2_{\odot}\simeq m_{2}^2-m_1^2$), 
 and atmospheric neutrino deficit is caused by  
 the mixing of the 2nd and the 3rd generations
 ($\theta_{atm}\simeq\theta_{23}$, 
 $\Delta m^2_{atm}\simeq m_3^2-m_2^2$).
We take 
 the LMA solution for the solar neutrino solution, 
 $\Delta m_{\odot}^2=(3-19)\times 10^{-5}$ eV$^2$ and 
 $\tan^2 2\theta_{\odot}=(0.25-0.65)$, from the recent 
 results including SK data\cite{SKsolar}\cite{SNO1}. 
Considering the results of the oscillation experiments,
 the hierarchical patterns of neutrino masses 
 are classified by the following three types: 
\begin{center}
\begin{tabular}{ccl}
A&:&$m_3\gg m_{1,2}$\\
B&:&$m_1\sim m_2\gg m_3$\\
C&:&$m_1 \sim m_2 \sim m_3$ \quad .
\end{tabular}
\end{center}
Taking into account of the mass squared differences, 
 $\Delta m_{\odot}^2$ and $\Delta m_{atm}^2$,
 the absolute 
 masses of the neutrino in the 
 leading are written by 
\begin{eqnarray}
\begin{minipage}{15.5cm}
\begin{flushleft}
\begin{tabular}{llcl}
\underline{Type A}&\\
&$m_1$&:&$0$\\
&$m_2$&:&$\sqrt{\Delta m_{\odot}^2}$\\
&$m_3$&:&$\sqrt{\Delta m_{atm}^2}$
\end{tabular}
\begin{tabular}{llcl}
\underline{Type B}&\\
&$m_1$&:&$\sqrt{\Delta m_{atm}^2}$\\
&$m_2$&:&$\sqrt{\Delta m_{atm}^2}
 +\frac{1}{2}\frac{\Delta m_{\odot}^2}{\sqrt{\Delta m_{atm}^2}}
$\\
&$m_3$&:&$0$
\end{tabular}
\begin{tabular}{llcl}
\underline{Type C}&\\
(1):\hspace{0.2cm}normal &\\
&$m_1$&:&$m_\nu$\\
&$m_2$&:&$m_\nu+\frac{1}{2}\frac{\Delta m_{\odot}^2}{m_\nu}$\\
&$m_3$&:&$m_\nu
+\frac{1}{2}\frac{\Delta m_{atm}^2}{m_\nu}$
\end{tabular}
\begin{tabular}{llcl}
\vspace*{0.2cm}\\
(2):\hspace{0.2cm}inverted \\
&$m_1$&:&$m_\nu+\frac{1}{2}\frac{\Delta m^2_{atm}}{m_{\nu }}$\\
&$m_2$&:&$m_\nu+\frac{1}{2}\frac{\Delta m^2_{atm}+\Delta
 m_{\odot}^2}{m_{\nu }}$\\
&$m_3$&:&$m_\nu $\quad
\end{tabular}
\end{flushleft}
\end{minipage}
\label{masstype}
\end{eqnarray}
in each type, respectively. 
Where $m_\nu$ in Type C is the scale of 
 the degenerated neutrino masses.
The recent neutrino oscilation experiment show\cite{SKsolar}{CHOOZ} 
\begin{eqnarray*}
\begin{array}{ll}
\Delta m_{atm}^2=3.2\times 10^{-3}\;\mathrm{eV^2}\;,&
\Delta m_{\odot}^2=6.9\times 10^{-5}\;\mathrm{eV^2}\;,\\
\sin^2 2\theta_{atm}=1.0\;,&
\tan^2\theta_{\odot}=3.6\times 10^{-1}\;,
\end{array}  
\end{eqnarray*}
and 
\begin{eqnarray*}
\sin^2 2\theta_{13}<0.1,
\end{eqnarray*}

In the zeroth order approximations, 
 $\displaystyle \cos \theta_{12}=\cos\theta_{23}={1}/{\sqrt{2}}$ and 
 $\sin\theta_{13}=0$, 
 we can obtain 
 the zeroth order form of the 
 MNS matrix as, 
\begin{eqnarray}
V^{(0)}=\left(
\begin{array}{ccc}
\frac{1}{\sqrt{2}}&\frac{1}{\sqrt{2}}&0\\
-\frac{1}{{2}}&\frac{1}{{2}}&\frac{1}{\sqrt{2}}\\
\frac{1}{2}&-\frac{1}{2}&\frac{1}{\sqrt{2}}
\end{array}
\right). 
\label{Vzero}
\end{eqnarray}
The neutrino mass matrix $M_\nu$ is 
 determined by ${}U$ and ${M}_{\nu}^{diag}$ 
 from Eqs.(\ref{MM})$\sim$(\ref{tildeM}). 
The zeroth order form of the neutrino mass 
 matrix is determined by the approximated MNS matrix, $V^{(0)}$, 
 according to the patterns of neutrino mass hierarchy,
 Types A$\sim$C.    
In Ref.\cite{Altarelli}, 
 the zeroth order forms of neutrino mass matrices 
 are shown when Majorana $CP$ phases are $0$ or 
 $\pi$, which are shown in Table 1. 
These mass matrices are 
 useful for the zeroth order approximations 
 of estimating the probability of $0\nu\beta\beta$ as follows.

 
The effective neutrino mass $\langle m\rangle$, which shows 
 the magnitude of $0\nu\beta\beta$ decay \cite{Ta}
 in Eq.(\ref{result}) 
 is defined by
\begin{eqnarray}
\langle m \rangle&=& |\sum_{i=1}^{3}U_{ei}^2 m_i |
=|\sum_{i=1}^{3}U_{ei} m_i U^{\mathrm{T}}_{ie} |
=|V_{e1}^2 m_1 e^{-i\phi_1}+V_{e2}^2 m_2 e^{-i\phi_2}+V_{e3}^2 m_3|,
\label{<m>}
\end{eqnarray}
where $i$ denotes the label of the mass 
 eigenstate ($i=1,2,3$). 
The value of $\langle m \rangle$ 
 is equal to 
the absolute value of $(1,1)$ component of ${M}_\nu$. 
Thus, Table 1 seems to suggest that 
 the forms of the neutrino mass matrix should be 
 B2 or C0 or C3,  
 in order to obtain the suitable 
 large magnitude 
 of $(1,1)$ component. 
However, it is too naive estimation. 
The previous paper Ref.\cite{HS} has shown 
 that C1 and C2 can also induce sizable 
 magnitude of $\langle m \rangle$ 
 by increasing the value of $m_\nu$.

How can the value of $m_\nu$ 
 be large? 
In fact there is the cosmological 
 upper bound for $m_\nu$, 
 $\Sigma = m_1 + m_2 + m_3 
 \leq 4.4$ eV \cite{cosmo}.  
This hot (neutrino) dark matter 
 constraint comes from 
 the CMB measurements and galaxy 
 cluster constructions.

Now let us 
 estimate the magnitude of $0\nu \beta\beta$ decay 
 in the degenerate neutrinos including 
 the cosmological constraint of 
 neutrino dark matter\cite{dark}. 
We will analyze this relation 
 including mass squared differences 
 of solar LMA solution and the value of $U_{e3}$ 
 in accordance with Majorana CP phases. 
At first, 
 we show the case of Type B2, 
 where the sign of $m_1$ is the same as that of 
 $m_2$. 
The approximation 
 in Table \ref{table:mass} shows 
 $\langle m \rangle=\mathcal{O}(\sqrt{\Delta m_{atm}^2})
 \sim 0.057$ eV. 
We can see 
 the value of 
 $\langle m \rangle$
 cannot be larger than 
 $0.06$ eV 
 even if we change the parameters of 
 $\phi_{1,2}$ and
 $U_{e3}$ in Type B \cite{HS}.  
The region where $\langle m \rangle > 0.05$ eV 
 only exists just around B2 in the 
 parameter space of $\phi_{1,2}$. 
This magnitude of $\langle m \rangle$ is 
 the edge of the allowed region of 
 experimental value of $97\%\;\mathrm{c.l.}$ 
 in Eq.(\ref{result}), and the value of $\Sigma$ 
 is 
 of order $2 \times \sqrt{\Delta m_{atm}^2}$, 
 which is much smaller than the cosmological 
 constraint, 4.4 eV. 
Thus, Type B neutrino 
 mass pattern automatically satisfies 
 the cosmological constraint. 
We would like to be concentrating on 
 the degenerate neutrino mass patterns 
 (Type C) from now on. \\

The neutrino masses are degenerate in Type C, and 
 we set the value of the degenerate mass as $m_\nu$.
In Ref.\cite{Altarelli}, 
 Type C mass hierarchy is classified to four subgroups, 
 C0, C1, C2 and C3, by relative signs of $m_1$, $m_2$ and $m_3$.
\begin{eqnarray*}
(\tilde{m}_1,\tilde{m}_2,\tilde{m}_3)=
\left\{
\begin{array}{rrrccccc}
m_\nu(&1,&1,&1)&\quad&(\phi_1,\phi_2)=(0,0)&\quad&(\mathrm{Type\;C0})\\
m_\nu(&-1,&1,&1)&&(\phi_1,\phi_2)=(0,\pi)&&(\mathrm{Type\;C1})\\
m_\nu(&1,&-1,&1)&&(\phi_1,\phi_2)=(\pi,0)&&(\mathrm{Type\;C2})\\
m_\nu(&-1,&-1,&1)&&(\phi_1,\phi_2)=(\pi,\pi)&&(\mathrm{Type\;C3})
\end{array}
\right.
\end{eqnarray*}
Seeing the zeroth order neutrino mass matrices 
 in Table \ref{table:mass}, 
 we suppose, naively, 
 only 
 Type C0 and C3 might explain $0\nu\beta\beta$
 decay experiments because the (1,1) element of 
 mass matrix is of $\mathcal{O}(m_\nu)$. 
However Type C1 and C2 
 cases can also explain Eq.(\ref{result}) 
 in the suitable large value of $m_{\nu} $\cite{HS}.

We show the value of $\langle m \rangle $ in cases of normal and inverted hierarchies of C0 $\sim $ C3
including $\Delta m^2_{\odot }$ and $\Delta m^2_{atm}$. In Type C0 and
C3, it is given by 
\begin{eqnarray*}
\langle m \rangle &=& \mid V^2_{e1}m_{\nu }+V^2_{e2}(m_{\nu
 }+\frac{\Delta m^2_{\odot }}{2m_{\nu }})\pm V^2_{e3}(m_{\nu
 }+\frac{\Delta m^2_{atm }}{2m_{m_{\nu }}})\mid , \hspace{2cm}({\rm normal})\\
&=& \mid V^2_{e1}(m_{\nu }+\frac{\Delta m^2_{atm }}{2m_{\nu }})+V^2_{e2}(m_{\nu
 }+\frac{\Delta m^2_{\odot }+\Delta m^2_{atm }}{2m_{\nu }})\pm V^2_{e3}m_{\nu
 }\mid .\hspace{0.2cm}({\rm inverted})\\
\end{eqnarray*}
By using the experimental values, $\mid \hspace{-0.1cm}V_{e3}\hspace{-0.1cm}\mid  \leq 0.1 $,
$0.48 \leq \mid \hspace{-1.5mm}V_{e2}\hspace{-1mm}\mid  \leq 0.63$, $ \Delta m_{\odot }=6.9\times 10^{-5} {\rm eV}^2$ and $ \Delta
m^2_{atm}=3.2\times 10^{-3}{\rm eV}^2$, 
we can see that the difference of normal and inverted hierarchies is
significant in the range,
\begin{eqnarray*}
m_{\nu }\leq {\frac{\Delta m^2_{atm }}{2m_{\nu }}}.
\end{eqnarray*}
It mean that we can neglect the difference of the normal and inverted
hierarchies in the region of $m^2_{\nu
}\gg 10^{-3 }$eV$^2$.
\\
On the other hand, in Type C1 and C2, the values of $\langle m \rangle $
are given by 
\begin{eqnarray*}
\langle m  \rangle 
&=&\mid (V^2_{e1}-V^2_{e2})m_{\nu }-V_{e2}^2\frac{\Delta m^2_{\odot
 }}{2m_{\nu }}\pm V^2_{e3}(m_{\nu }+\frac{\Delta m^2_{atm }}{2m_{\nu
 }})\mid ,\hspace{2cm}({\rm normal})\\ 
\langle m  \rangle 
&=& \mid (V^2_{e1}-V^2_{e2})m_{\nu }-V_{e2}^2\frac{\Delta m^2_{\odot
 }}{2m_{\nu }}\pm V^2_{e3}m_{\nu }-\frac{\Delta m^2_{atm }}{2m_{\nu
 }}(V^2_{e2 }-V^2_{e1})\mid .\hspace{0.2cm}({\rm inverted })
\end{eqnarray*}
\\
They suggest the difference of normal and inverted
hierarchies  is significant in the range of
\begin{eqnarray*}
\mid (V^2_{e1}-V^2_{e2})m_{\nu }-V^2_{e2}\frac{\Delta m^2_{\odot
 }}{2m_{\nu }}\pm V^2_{e3}m_{\nu }\mid \leq \mid  V^2_{e3}\frac{\Delta
 m^2_{atm }}{2m_{\nu }} \mid ,\\
\mid (V^2_{e1}-V^2_{e2})m_{\nu }-V^2_{e2}\frac{\Delta m^2_{\odot
 }}{2m_{\nu }}\pm V^2_{e3}m_{\nu }\mid \leq \mid \frac{\Delta m^2_{atm
 }}{2m_{\nu }}(V_{e2}^2-V_{e3}^2)\mid . 
\end{eqnarray*}
Thus, we can neglect the difference between the normal and inverted
hierarchies in $m_{\nu }\gg 10^{-3}$eV$^2$. 
Above discussions mean the value of $\langle m \rangle $ does not depend on
whether neutrino mass is normal or inverted hierarchies in the range of
$m_{\nu}^2 \geq 10^{-2} $eV$^2$. We can see it explicitly in Fig.1 (normal
hierarchy case) and Fig.2 (inverted hierarchy case).  
\setlength{\unitlength}{1cm}
\begin{figure}[t]
\begin{center}
\begin{picture}(14,18)
\put(4,16){(1):\hspace{0.2cm}normal hierarchy case}
\put(0,11){\includegraphics[scale=.5]{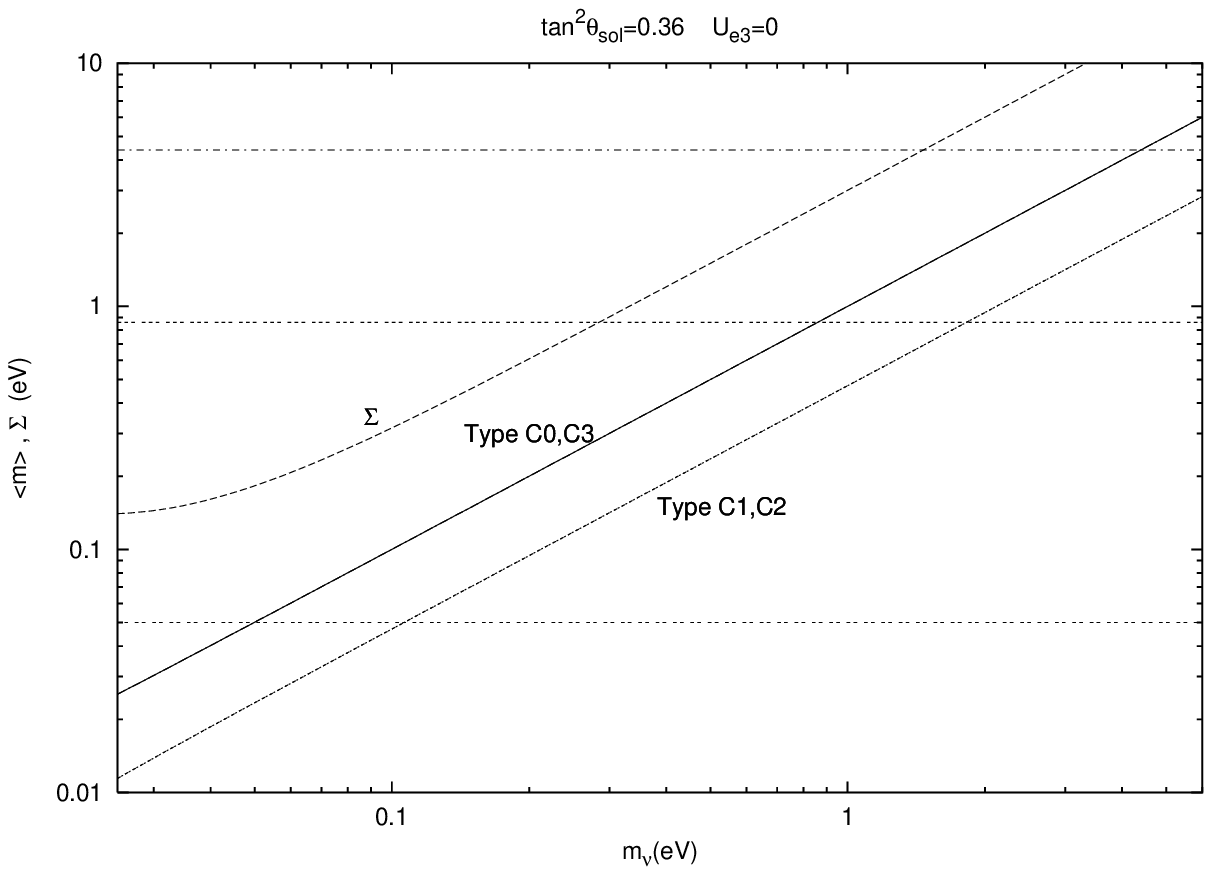}}
\put(7,11){\includegraphics[scale=.5]{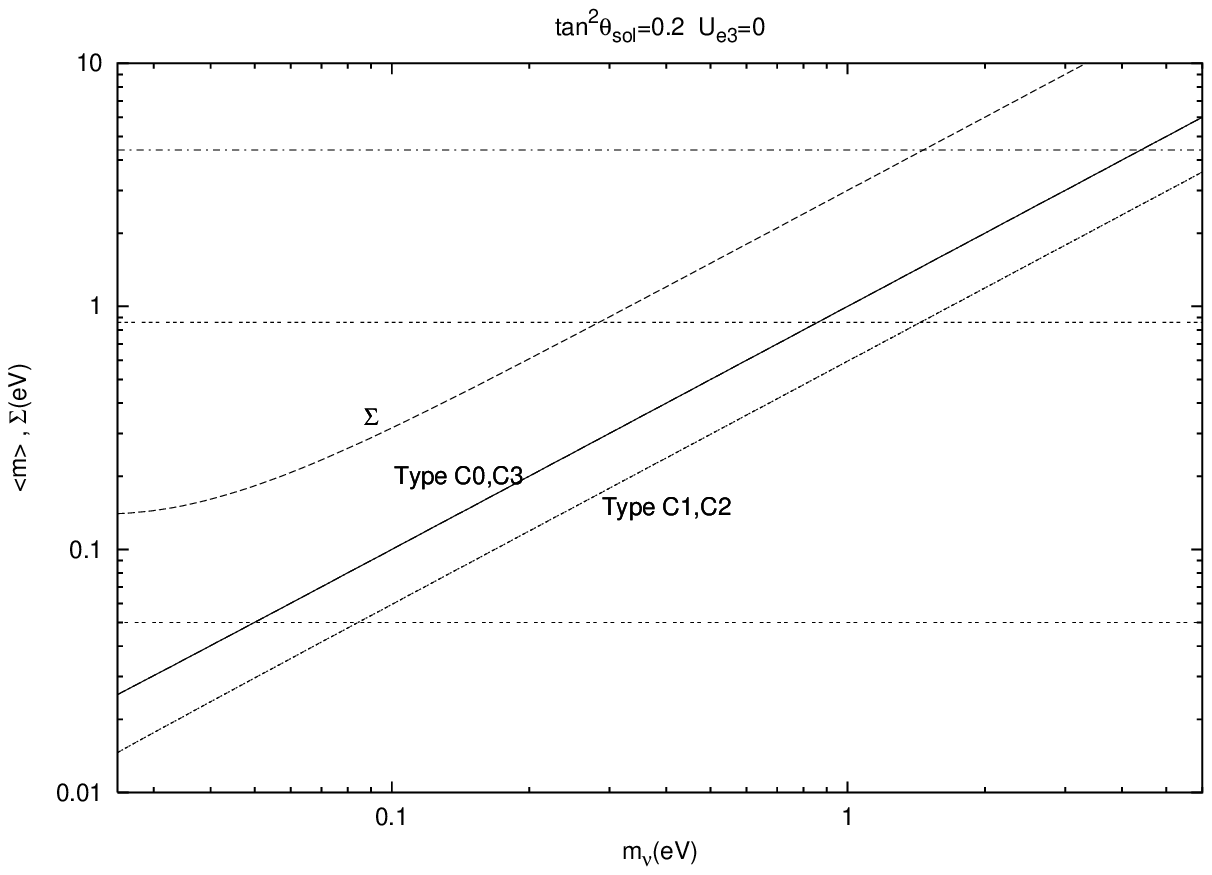}}
\put(3,10.5){(1.a)} \put(10,10.5){(1.b)}
\put(0,5.5){\includegraphics[scale=.5]{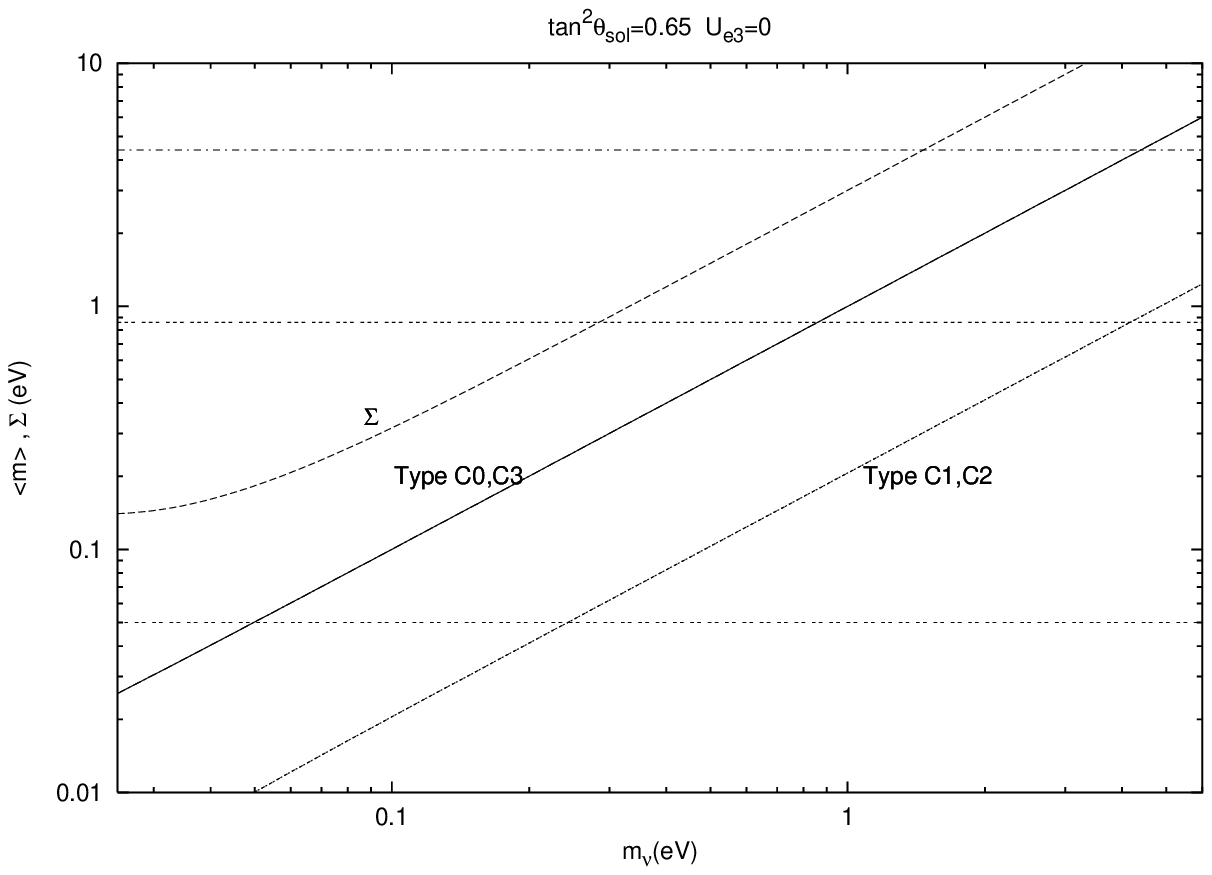}}
\put(7,5.5){\includegraphics[scale=.5]{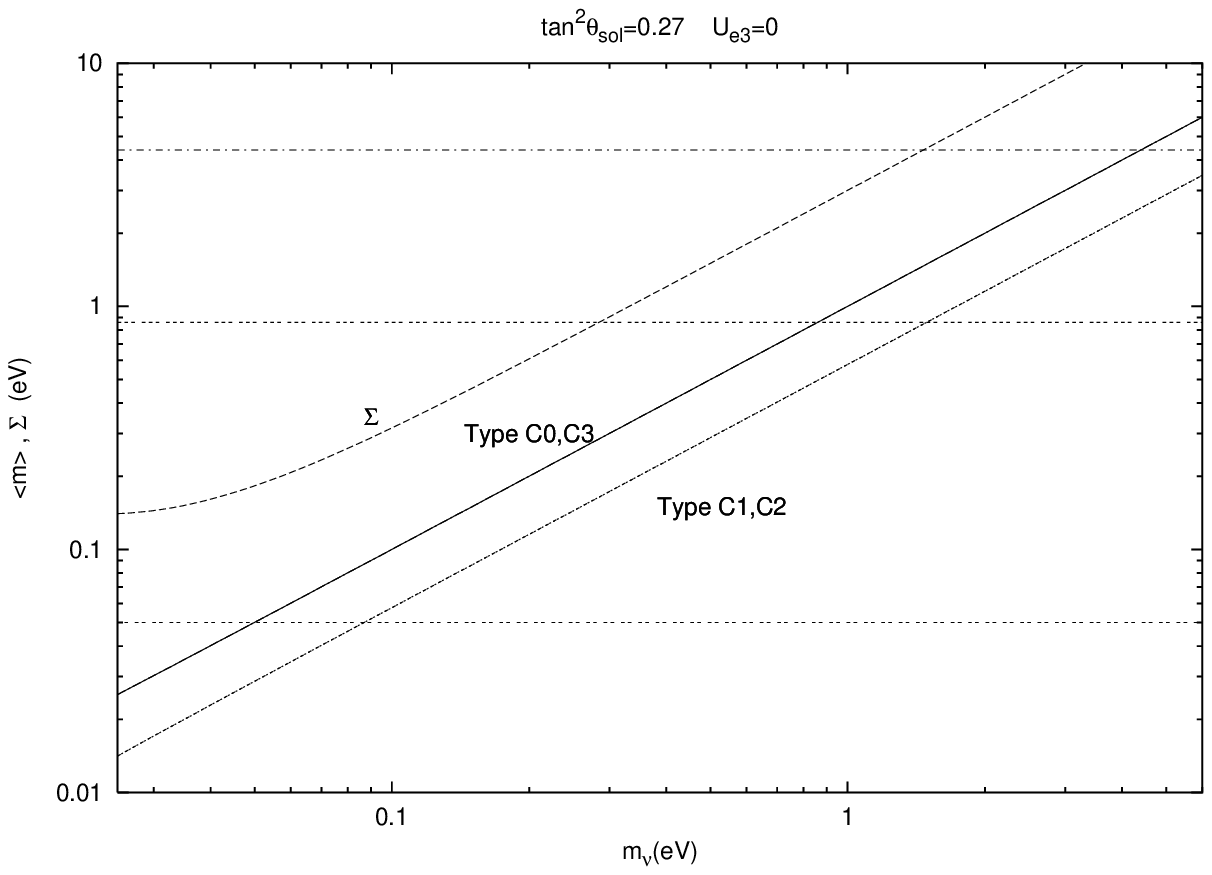}}
\put(3,5){(1.c)} \put(10,5){(1.d)}
\put(0,0){\includegraphics[scale=0.5]{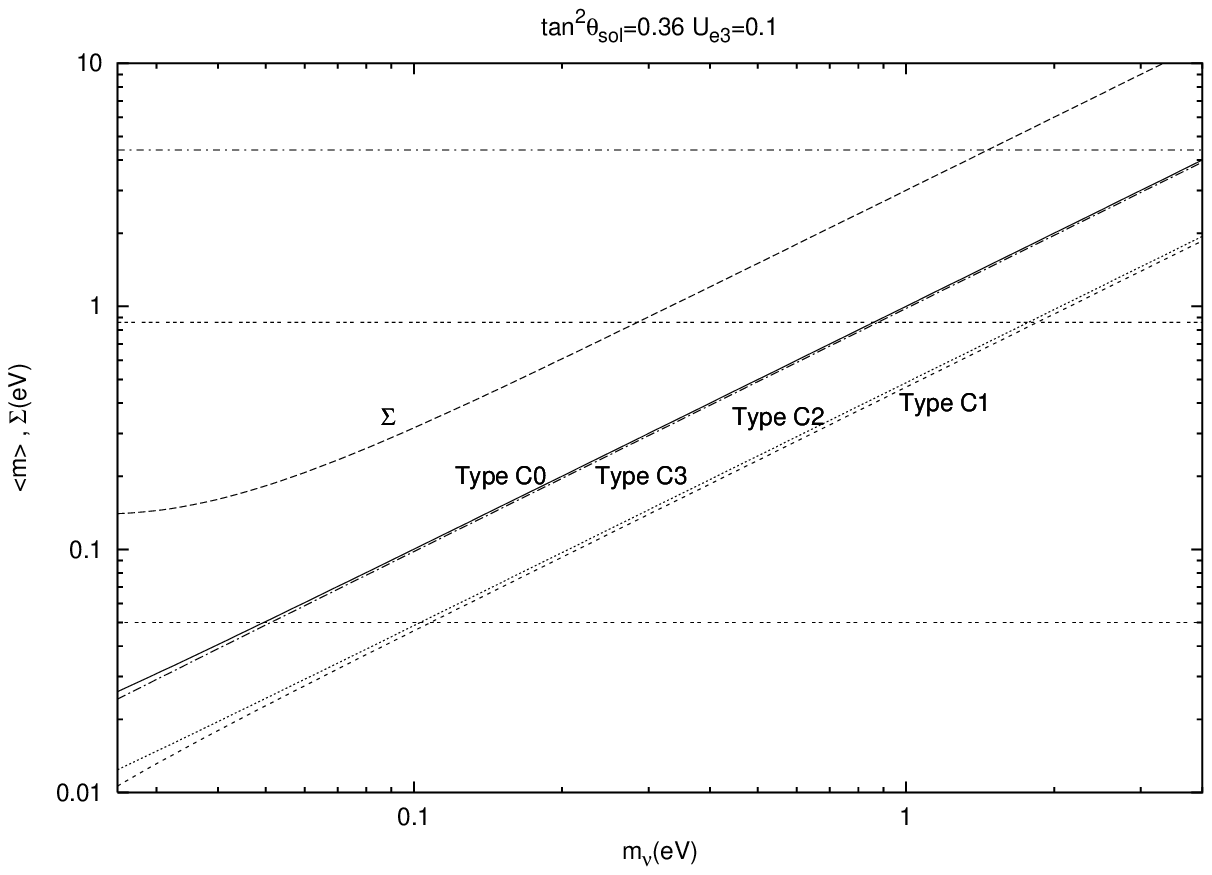}}
\put(3,-0.5){(1.e)}
\end{picture}
\end{center}
\caption{
The values of $\langle m \rangle$ and $\Sigma$
 with $\tan^2 \theta_{\odot}=0.34$ (the LMA center value\cite{SNO1}) in 
 (1.a), $\tan^2 \theta_{\odot}= 0.1$  in (1.b), and 
 $\tan^2 \theta_{\odot}= 0.5$  in (1.c).
 Other values are taken as the zeroth order values as $U_{e3}=0$
 and $\theta_{atm}=\pi/4$ in (1.a)$\sim $(1.c) 
We show
the values of $\langle m \rangle$ and $\Sigma$
 with $\tan^2 \theta_{\odot}=0.27$,       
 $U_{e3}=0$ in (1.d) and $\tan^2 \theta_{\odot}=0.34$ (the LMA center
 values), $U_{e3}$=0.1
 in (1.e).        
The allowed region are
$0.05 \;\mathrm{eV}
< \langle m \rangle <
0.86\;\mathrm{eV}$ from $0\nu\beta\beta$ and
$\Sigma \leq 4.4\;\mathrm{eV}$ from cosmology.}
\label{FIGURE}
\end{figure}

\vspace*{0.5cm}
We stress here that the maximal (minimum) value of $\langle m \rangle$ is 
 given by C0 (C1) line 
 independently of Majorana phases. It is because the relation of $\hspace{-1mm}\mid \hspace{-0.1cm}V_{e1}^2m_1-V_{e2}^2m_2\hspace{-0.1cm} \mid >\mid  \hspace{-0.1cm}V_{e3}^3m_3\hspace{-0.1cm}\mid $ is
 always satisfied in Eq.(7) when $m_{\nu } > 2.8\times 10^{-2}$eV
, $\mid
 \hspace{-0.1cm}V_{e3} \hspace{-0.1cm}\mid \leq 0.1 $ and $0.30
 \leq \mid\hspace{-0.1cm} V_{e2}\hspace{-0.1cm}\mid \leq 0.58 $.
\\
 Let us show the normal hierarchy case at first.  
Figures (1.a)$\sim $(1.c) show the 
 values of $\langle m \rangle$ and $\Sigma$ 
for the value of $m_{\nu } $ with $U_{e3}=0$. When $U_{e3}=0$, C3 (C2) line falls on C0 (C1) line. Then, the maximal (minimum) value of 
 $\langle m \rangle$ is induced by 
 the cases C0, C3 (C1, C2) in the given 
 value of $m_\nu$. 
In Fig.(1.a) we take the center value of 
 LMA solution as $\tan^2 \theta_\odot =0.36$. 
We can see that $0 \nu \beta \beta$ decay 
 constraint (Eq.(\ref{result})) is 
 severer than the cosmological constraint, 
 $\Sigma \leq 4.4$ eV, in Type C0 and C3. 
On the other hand, 
 the cosmological constraint 
 is severer than $0\nu \beta\beta$ 
 result in cases of C1 and C2. 
Figure (1.b) ((1.c)) shows the case of 
 $\tan^2 \theta_\odot =0.2$ $(0.65)$. 
In this case 
 the line of minimum value $\langle m \rangle$,
 C1, C2, 
 is lowered (lifted) since the cancellation between 
 $m_1$ and $m_2$ in Eq.(\ref{<m>}) is (not) enhanced.
Other lines, C0, C3, and $\Sigma$, are 
 not changed from Fig.(1.a). 
The case of $\tan^2 \theta_\odot =0.65$ 
 shows $0\nu \beta\beta$ result is 
 severer than the cosmological constraint as
 shown in Fig.(1.c). 
\setlength{\unitlength}{1cm}
\begin{figure}[t]
\begin{center}
\begin{picture}(14,18)
\put(4,16){(2):\hspace{0.2cm}inverted hierarchy case}
\put(0,11){\includegraphics[scale=0.5]{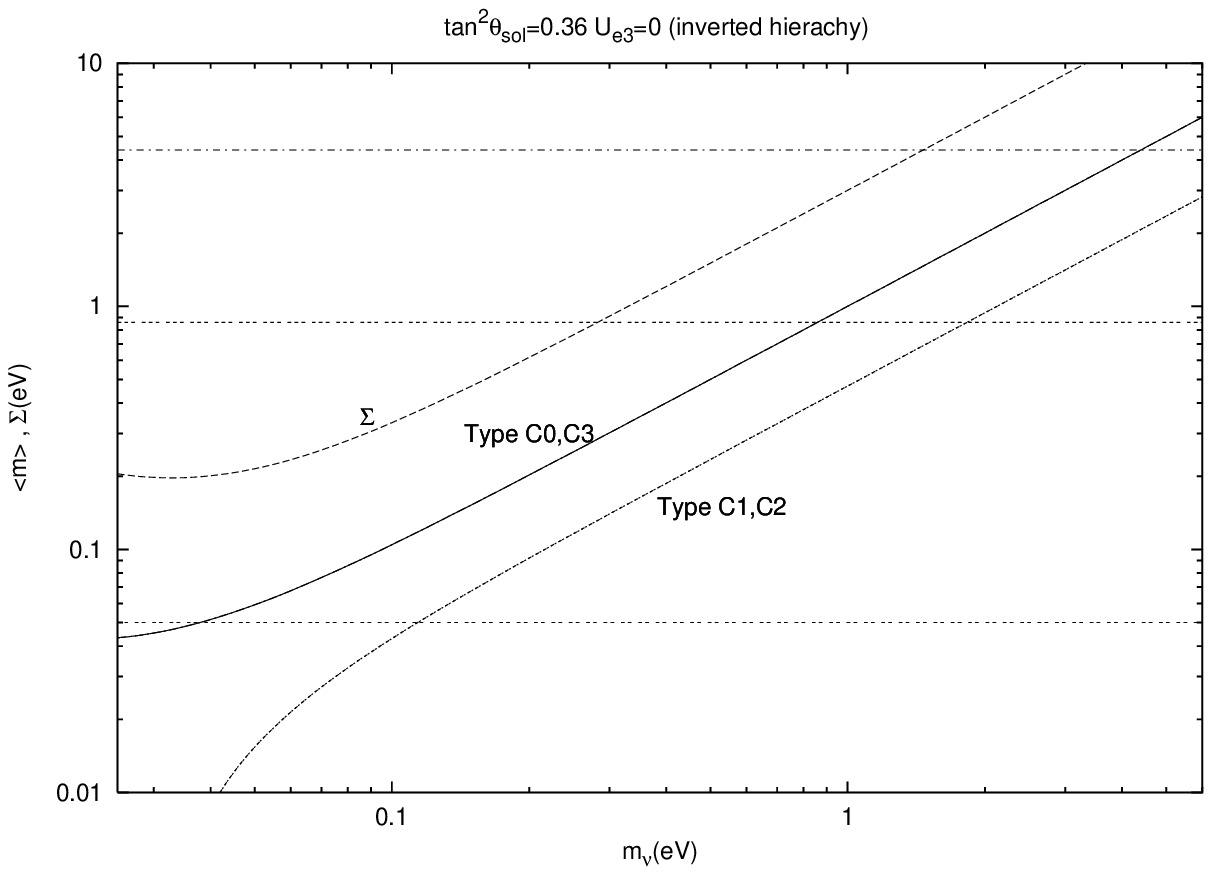}}
\put(7,11){\includegraphics[scale=0.5]{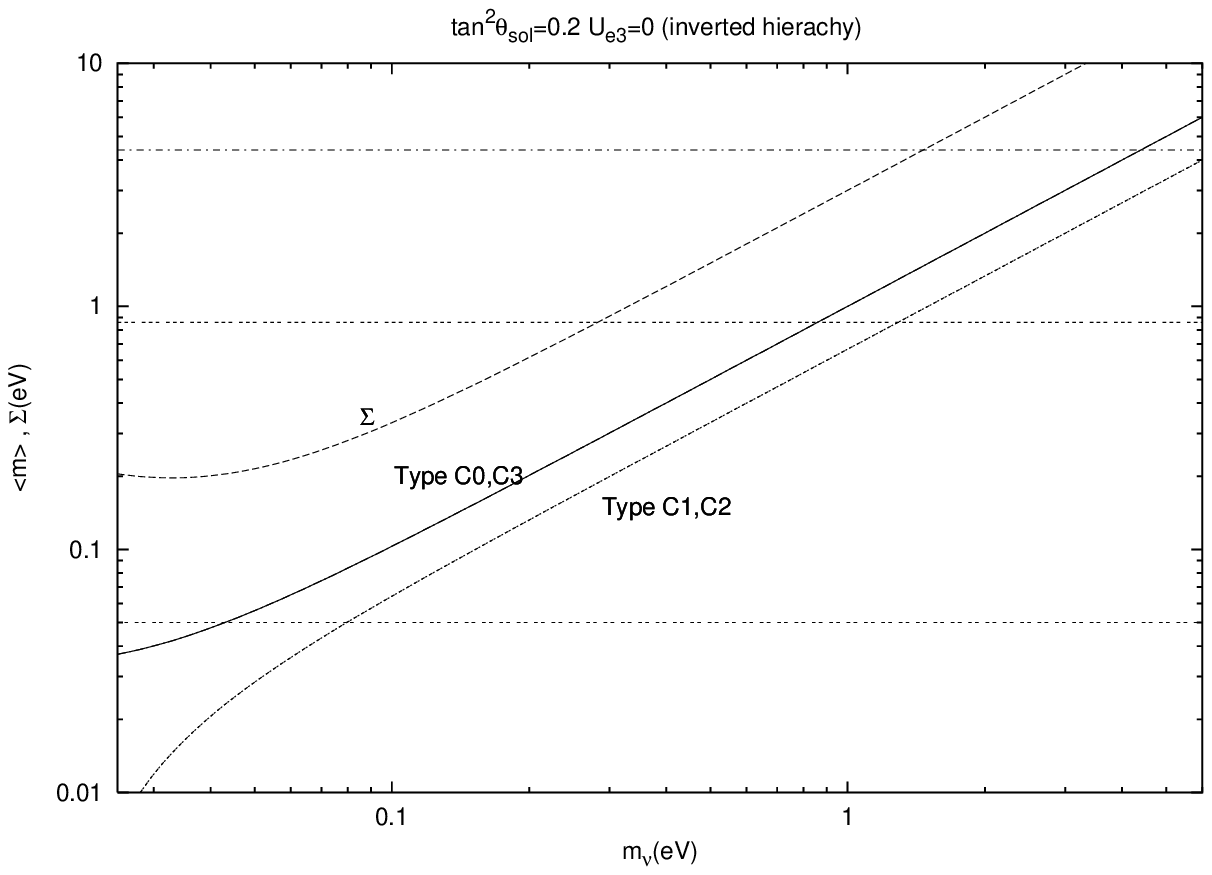}}
\put(3,10.5){(2.a)}\put(10,10.5){(2.b)}
\put(0,5.5){\includegraphics[scale=0.5]{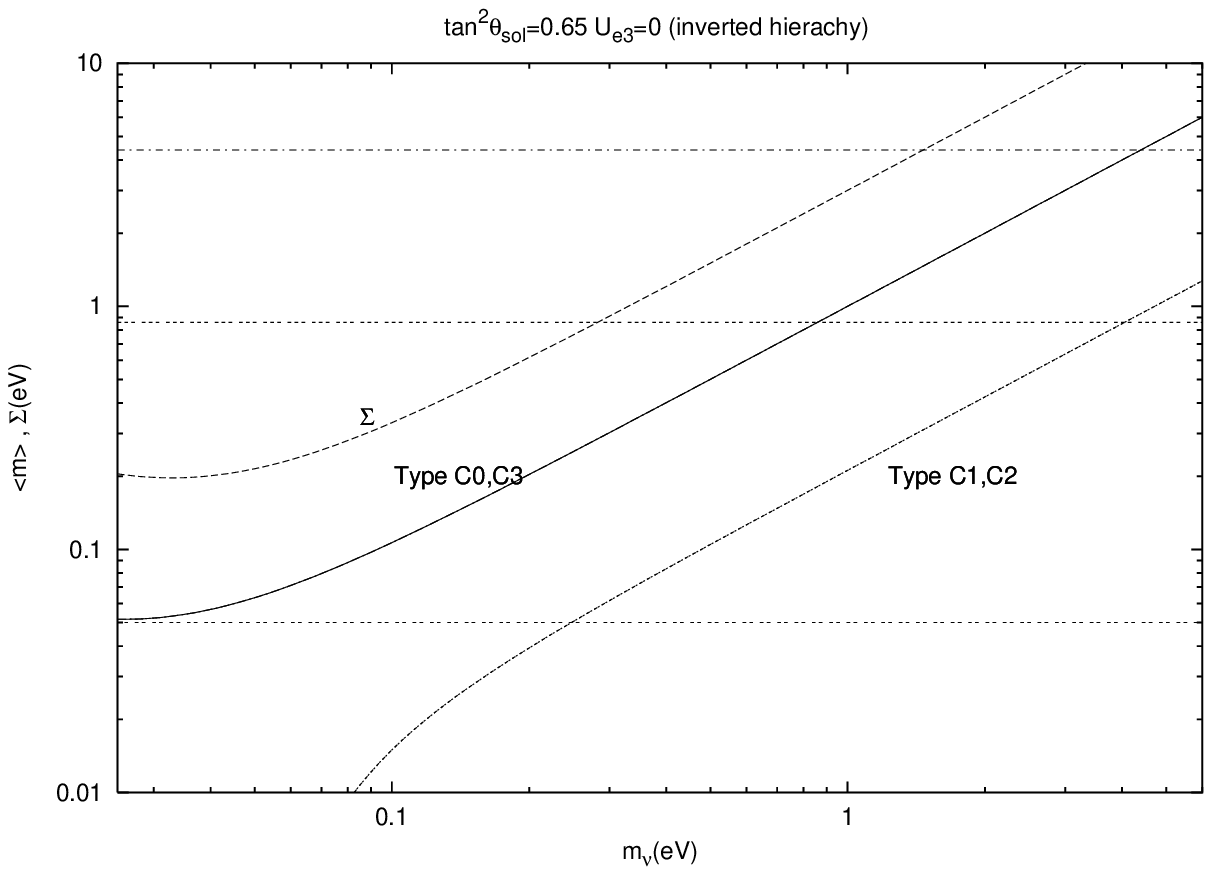}}
\put(7,5.5){\includegraphics[scale=0.5]{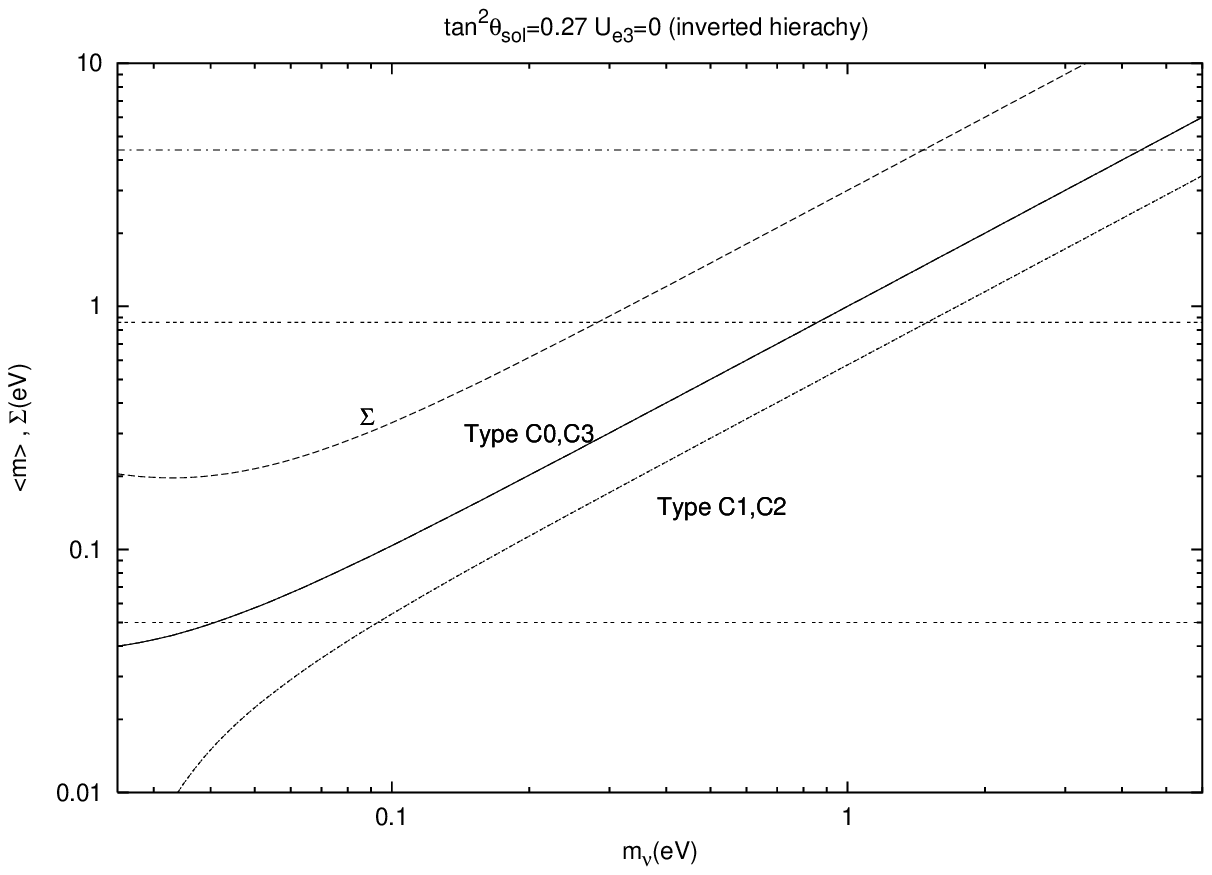}}
\put(3,5){(2.c)}\put(10,5){(2.d)}
\put(0,0){\includegraphics[scale=0.5]{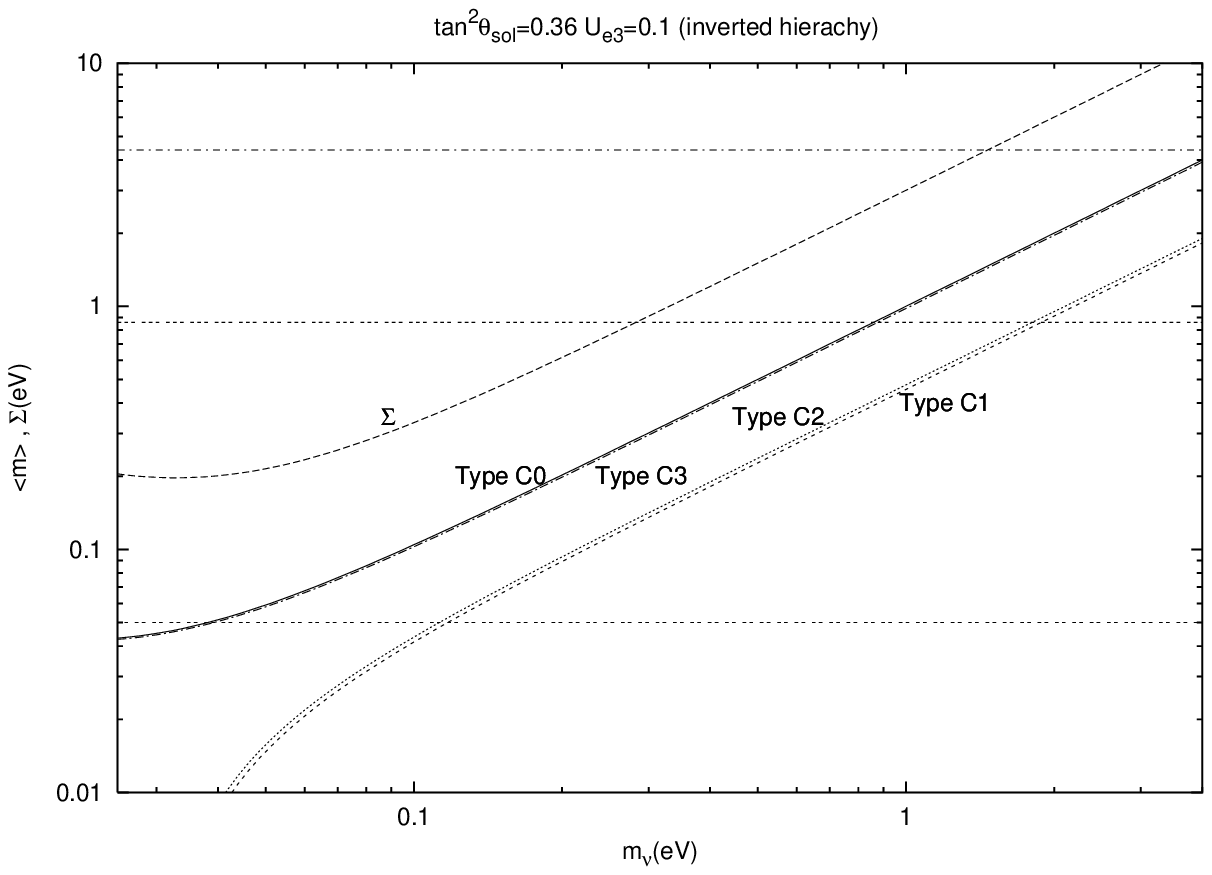}}
\put(3,-0,5){(2.e)}
\end{picture}
\end{center}
\caption{
The values of $\langle m \rangle$ and $\Sigma$
 with $\tan^2 \theta_{\odot}=0.36$ (the LMA center value\cite{SNO1}) in 
 (2.a), $\tan^2 \theta_{\odot}= 0.2$  in (2.b), and 
 $\tan^2 \theta_{\odot}= 0.65$  in (2.c).
 Other values are taken as the zeroth order values as $U_{e3}=0$
 and $\theta_{atm}=\pi/4$ in (2.a)$\sim $(2.c) 
We show
the values of $\langle m \rangle$ and $\Sigma$
 with $\tan^2 \theta_{\odot}=0.27$, 
 $U_{e3}=0$ in (2.d) and $\tan^2 \theta_{\odot}=0.36$ (the LMA center value), $U_{e3}$=0.1
 in (2.e).      
The allowed region are
$0.05 \;\mathrm{eV}
< \langle m \rangle <
0.86\;\mathrm{eV}$ from $0\nu\beta\beta$ and
$\Sigma \leq 4.4\;\mathrm{eV}$ from cosmology.}
\label{FIGURE}
\end{figure}
Figure (1.d) show the case of $\tan^2\theta _{\odot }=0.27$ and
 $U_{e3}=0$. In this case, the cosmological constraint is the same as  $0\nu \beta \beta $
 decay constraint in Type C1 and C2. However, the cosmological
 constraint is severer than $0 \nu \beta \beta $ decay constraint  in
 Type C0 and C3.   
Figure (1.e) shows the case of $U_{e3}=0.1$ 
 with LMA center value. This case split C0 from C3, and C1 from
 C2. However, this effect is not so large.   
 All figures show that tritium $\beta$ decay constraint, 
 $m_\nu < 3$ eV, is less severer constraint. 

 Figure 2 show the inverted hierarchy cases.
We finds almost same  results as the normal hierarchy
case in the range of Eq.(1), since where $m_{\nu }^2\geq 10^{-2}(\gg 10^{-3})$eV$^2$.



\hspace{1cm}

A recent paper\cite{db} 
 announces the evidence of 
 $0\nu\beta\beta$ decay, and  
 the value of $\langle m \rangle$ 
 is large as Eq.(\ref{result}). 
This means that neutrinos are Majorana particles and 
 their mass hierarchy is forced 
 to Type B or 
 Type C
 patterns in the diagonal basis of
 charged lepton mass matrix. 
In this paper we have estimated 
 the magnitude of $0\nu \beta\beta$ decay 
 in the degenerate neutrinos including 
 the cosmological constraint of 
 neutrino dark matter, since 
 the Type B mass hierarchy pattern
 always satisfies the cosmological 
 constraint.  
The absolute value of neutrino mass 
 is constrained by $0\nu \beta\beta$ result or 
 cosmology depending on Majorana $CP$ phases
 of neutrino masses and solar mixing angle. 
C0 and C3 cases are constrained from 
 $0\nu \beta\beta$ result. 
On the other hand, the constraint of neutrino absolute mass with $\Delta
 m^2_{atm}=3.2\times 10^{-3}$ and $\Delta m^2_{\odot }=6.9\times
 10^{-5}$ is gived by
 the following. 
\begin{center}
\begin{tabular}{|c|c|}
\hline 
 solar mixing angle & constranint \\
\hline
$0.27<\tan^2\theta _{\odot }\leq 0.65 $& cosmology 
  \\
$\tan^2\theta _{\odot }=0.27$ & cosmology and $0\nu \beta \beta$ decay \\
 $0.2\leq \tan^2\theta _{\odot} < 0.27$ &  $0\nu \beta \beta $ decay\\
\hline
\end{tabular}
\end{center}

{\small
\begin{table}[b]
\begin{minipage}{8.3cm}
\hspace{-1.7cm}
\begin{tabular}{|c|c|c|}
\hline     & &Neutrino  \\
           & $\widetilde{M}_\nu$  &mass matrix        \\
\hline & & \\
 A &\small $diag.(0,0,1)$ &
$\left[ 
\begin{array}{ccc} 
 0&0&0\\
 0&\um&\um\\
 0&\um&\um
\end{array}
\right]$ 
 \\ & &  \\
\hline & &  \\
 B1 &\small$diag.(1,-1,0)$&
\small$\left[
\begin{array}{ccc}
 0&-\sq&\sq\\
 -\sq&0&0\\
 \sq&0&0
\end{array}
\right]$ 
 \\ & &  \\
\hline & &  \\ B2 &\small$diag.(1,1,0)$&
$\left[
\begin{array}{ccc}
 1&0&0\\
 0&\um&-\um\\
 0&-\um&\um
\end{array}
\right]$ 
 \\ & &  \\
\hline 
\end{tabular}
\vspace{1.7cm}
\end{minipage}
\begin{minipage}{8cm}
\hspace{-.7cm}
\begin{tabular}{|c|c|c|}\hline
&&\\
C0 &\small $diag.(1,1,1)$&
$\left[
\begin{array}{ccc}
 1&0&0\\
 0&1&0\\
 0&0&1
\end{array}
\right]$ 
 \\ & &  \\
\hline & & \\ C1 &\small $diag.(-1,1,1)$&
\small$\left[
\begin{array}{ccc}
 0&\sq&-\sq\\
 \sq&\um&\um\\
 -\sq&\um&-\um
\end{array}
\right]$ 
 \\ & & \\
\hline & & \\ C2 &\small$diag.(1,-1,1)$&
\small$\left[
\begin{array}{ccc}
 0&-\sq&\sq\\
 -\sq&\um&\um\\
 \sq&\um&\um
\end{array}
\right]$ 
 \\ & & \\
\hline & & \\ C3 &\small $diag.(-1,-1,1)$&
$\left[
\begin{array}{ccc}
-1&0&0\\
 0&0&1\\
 0&1&0
\end{array}
\right]$ 
 \\ & & \\
\hline
\end{tabular}
\end{minipage}
\caption
{The zeroth order neutrino mass matrices. 
In Type A and B, 
the eigenvalues of $\widetilde{M}_\nu$ and the neutrino mass matrices 
are normalized by $\sqrt{\Delta m_{atm}^2}.$
In Type C,
they are normalized by $m_\nu$.
}
\label{table:mass}
\end{table}
}

The results of MAP satellite will make $\Sigma < 0.5$ eV, 
 which will suggest more severer bound from the 
 cosmology\cite{MAP}.

\section*{Acknowledgment}
This work is supported in part by the Grant-in-Aid for Science Research,
 Ministry of Education, Science and Culture, Japan 
 (No. 14039207, No. 14046208, No. 14740164).



\begin{thebibliography}{99}

\bibitem{Kamiokande}
Y.~Fukuda {\it et al.}  [Kamiokande Collaboration],
Phys.\ Rev.\ Lett.\ {\bf 77} (1996) 1683.

\bibitem {SKatm}
Y.~Fukuda {\it et al.}  [Kamiokande Collaboration],
Phys.\ Lett.\ B {\bf 335} (1994) 237;
\\
Y.~Fukuda {\it et al.}  [Super-Kamiokande Collaboration],
Phys.\ Rev.\ Lett.\ {\bf 81} (1998) 1562;
\\
T.~Kajita  [Super-Kamiokande Collaboration], 
in {\it Neutrino Physics and Astrophysics},
Proceedings of the XVIIIth International Conference on Neutrino
Physics and Astrophysics (Neutrino '98), June 4-9, 1998, Takayama,
Japan, edited by Y. Suzuki and Y. Totsuka,
(Elsevier Science B.V., Amsterdam, 1999) page 123;
Nucl.\ Phys.\ Proc.\ Suppl.\ {\bf 77}, 123 (1999).





\bibitem{SKsolar}
Y.~Fukuda {\it et al.}  [Super-Kamiokande Collaboration],
\\
Phys.\ Rev.\ Lett.\ {\bf 81} (1998) 1158;
Erratum-ibid.\ {\bf 81} (1998) 4279;
\\
Phys.\ Rev.\ Lett.\ {\bf 82} (1999) 2430;
\\
Phys.\ Rev.\ Lett.\ {\bf 82} (1999) 1810;
 S.~Fukuda {\it et al.}  [Super-Kamiokande Collaboration],
Phys.\ Lett.\ B {\bf 539}, 179 (2002), arXiv:hep-ex/0205075.


\bibitem{SNO}

Q.~R.~Ahmad {\it et al.}  [SNO Collaboration],
Phys.\ Rev.\ Lett.\  {\bf 87}, 071301 (2001),
 arXiv:nucl-ex/0106015.




\bibitem{Cl-Homestake}
K.~Lande {\it et al.},
Astrophys.\ J.\  {\bf 496} (1998) 505.


\bibitem{Ga-Gallex-and-GNO}
V.~N.~Gavrin  [SAGE Collaboration],
Nucl.\ Phys.\ Proc.\ Suppl.\  {\bf 91} (2001) 36;
\\
E.~Bellotti,
Nucl.\ Phys.\ Proc.\ Suppl.\  {\bf 91} (2001) 44.


\bibitem{post-SNO-analysis}
V.~D.~Barger, D.~Marfatia and K.~Whisnant,
Phys.\ Rev.\ Lett.\  {\bf 88}, 011302 (2002);
\\
G.~L.~Fogli, E.~Lisi, D.~Montanino and A.~Palazzo,
Neutrino Observatory results,''
Phys.\ Rev.\ D {\bf 64}, 093007 (2001);
\\
J.~N.~Bahcall, M.~C.~Gonzalez-Garcia and C.~Pena-Garay,
measurement,''
JHEP {\bf 0108}, 014 (2001);
\\
A.~Bandyopadhyay, S.~Choubey, S.~Goswami and K.~Kar,
Phys.\ Lett.\ B {\bf 519}, 83 (2001).


\bibitem{SNO1}
Q.~R.~Ahmad {\it et al.}  [SNO Collaboration],
arXiv:nucl-ex/0204008.

Q.~R.~Ahmad {\it et al.}  [SNO Collaboration],
arXiv:nucl-ex/0204009.

\bibitem{db}
H.~V.~Klapdor-Kleingrothaus, A.~Dietz, H.~L.~Harney and I.~V.~
Krivosheina,
Mod.\ Phys.\ Lett.\ A {\bf 16}, 2409 (2001)


\bibitem{bb}
C.~E.~Aalseth {\it et al.},
arXiv:hep-ex/0202018;

F.~Feruglio, A.~Strumia and F.~Vissani,
arXiv:hep-ph/0201291;

H.~V.~Klapdor-Kleingrothaus,
arXiv:hep-ph/0205228.




\bibitem{OTHERS}
H.V. Klapdor-Kleingrothaus and U. Sarkar, Mod. Phys. Lett. {\bf A16}, 
2469 (2001), hep-ph/0201224;
V. Barger, S.L. Glashow, D. Marfatia, and K. Whisnant, hep-ph/0201262;
S.~F.~King,
 arXiv:hep-ph/0204360;
 Z.~z.~Xing,
 arXiv:hep-ph/0205032; 
 D.~Falcone,
 arXiv:hep-ph/0204335;
 X.~J.~Bi and Y.~B.~Dai,
 arXiv:hep-ph/0204317;
 F.~Simkovic, P.~Domin and A.~Faessler,
 arXiv:hep-ph/0204278;
 K.~Matsuda, T.~Kikuchi, T.~Fukuyama and H.~Nishiura,
 arXiv:hep-ph/0204254;
 D.~V.~Ahluwalia and M.~Kirchbach,
 arXiv:hep-ph/0204144;
 K.~W.~Edwards {\it et al.}  [CLEO Collaboration],
 arXiv:hep-ex/0204017;
 H.~B.~Nielsen and Y.~Takanishi,
 arXiv:hep-ph/0204027;
 N.~N.~Singh, N.~N.~Singh and M.~Patgiri,
 arXiv:hep-ph/0204021;
 G.~Barenboim, J.~F.~Beacom, L.~Borissov and B.~Kayser,
 arXiv:hep-ph/0203261;
 H.~J.~He, D.~A.~Dicus and J.~N.~Ng,
 arXiv:hep-ph/0203237;
 W.~Rodejohann,
 arXiv:hep-ph/0203214;
 Z.~Fodor, S.~D.~Katz and A.~Ringwald,
 arXiv:hep-ph/0203198;
 S.~Pakvasa and P.~Roy,
 arXiv:hep-ph/0203188;
 H.~B.~Nielsen and Y.~Takanishi,
 arXiv:hep-ph/0203147;
 Y.~Koide and A.~Ghosal,
 arXiv:hep-ph/0203113;
 M.~Y.~Cheng and K.~m.~Cheung,
 arXiv:hep-ph/0203051;
 S.~R.~Elliott and P.~Vogel,
 arXiv:hep-ph/0202264;
 M.~Frigerio and A.~Y.~Smirnov,
 arXiv:hep-ph/0202247;
 M.~Fujii, K.~Hamaguchi and T.~Yanagida,
 arXiv:hep-ph/0202210;
 H.~Nishiura, K.~Matsuda, T.~Kikuchi and T.~Fukuyama,
Phys.\ Rev.\ D {\bf 65}, 097301 (2002), 
 arXiv:hep-ph/0202189;
 C.~Giunti and M.~Laveder,
 arXiv:hep-ph/0202152;
 G.~Bhattacharyya, S.~Goswami and A.~Raychaudhuri,
 arXiv:hep-ph/0202147;
 N.~Haba and T.~Suzuki,
 arXiv:hep-ph/0202143;
 I.~Jack, D.~R.~Jones and R.~Wild,
 arXiv:hep-ph/0202101;
 Z.~z.~Xing,
Phys.\ Rev.\ D {\bf 65}, 077302 (2002),
 arXiv:hep-ph/0202034;
 G.~C.~Branco, R.~Gonzalez Felipe, F.~R.~Joaquim and M.~N.~Rebelo,
 arXiv:hep-ph/0202030;
 H.~Minakata and H.~Sugiyama,
Phys.\ Lett.\ B {\bf 532}, 275 (2002),
 arXiv:hep-ph/0202003;
 T.~Hambye,
 arXiv:hep-ph/0201307;
 Y.~Uehara,
 arXiv:hep-ph/0201277;
 H.~V.~Klapdor-Kleingrothaus and U.~Sarkar,
 arXiv:hep-ph/0201226;
 E.~Ma,
Mod.\ Phys.\ Lett.\ A {\bf 17}, 289 (2002),
 arXiv:hep-ph/0201225.



\bibitem{dark}
V.~Barger, S.~L.~Glashow, D.~Marfatia and K.~Whisnant,
arXiv:hep-ph/0201262.



\bibitem{seesaw}
T.~Yanagida,
{\it ``Horizontal Symmetry And Masses Of Neutrinos''},
Prog.\ Theor.\ Phys.\  {\bf 64} (1980) 1103,
and in Proceedings of the 
{\it ``Workshop on the Unified Theory and the Baryon Number in the
  Universe''}, Tsukuba, Japan, Feb 13-14, 1979, 
  Eds. O.~Sawada and A.~Sugamoto, KEK report KEK-79-18, p. 95;
\\
M.~Gell-Mann, P.~Ramond and R.~Slansky,
{\it in} ``Supergravity''
(North-Holland, Amsterdam, 1979)
{\it eds.} D.Z. Freedman and P. van Nieuwenhuizen,
Print-80-0576 (CERN).


\bibitem{HS}
N.~Haba and T.~Suzuki,
arXiv:hep-ph/0202143.




%

\bibitem{Altarelli}
G.~Altarelli and F.~Feruglio,
Phys.\ Rept.\  {\bf 320}, 295 (1999).

\bibitem{cosmo}
X.~Wang, M.~Tegmark and M.~Zaldarriaga,
arXiv:astro-ph/0105091.
\bibitem {MNS}Z. Maki, M. Nakagawa, and S. Sakata, Prog. Theor. Phys.
     {\bf 28} (1962) 870.

\bibitem{CHOOZ}
M.~Apollonio {\it et al.}  [CHOOZ Collaboration],
Phys.\ Lett.\ B {\bf 466}, 415 (1999).

\bibitem{Ta}
S.~Pascoli and S.~T.~Petcov,
 arXiv:hep-ph/0205022;
 S.~Pascoli and S.~T.~Petcov,
 arXiv:hep-ph/0111203;
 S.~M.~Bilenky, S.~Pascoli and S.~T.~Petcov,
Phys.\ Rev.\ D {\bf 64}, 113003 (2001),
 arXiv:hep-ph/0104218;
 S.~M.~Bilenky, S.~Pascoli and S.~T.~Petcov,
Phys.\ Rev.\ D {\bf 64}, 053010 (2001),
 arXiv:hep-ph/0102265;
 S.~M.~Bilenky, C.~Giunti, C.~W.~Kim and S.~T.~Petcov,
Phys.\ Rev.\ D {\bf 54}, 4432 (1996),
 arXiv:hep-ph/9604364;
 S.~M.~Bilenkii, J.~Hosek and S.~T.~Petcov,
Phys.\ Lett.\ B {\bf 94}, 495 (1980);
 S.~M.~Bilenky, C.~Giunti, W.~Grimus, B.~Kayser and S.~T.~Petcov,
Phys.\ Lett.\ B {\bf 465}, 193 (1999),
 arXiv:hep-ph/9907234;
 S.~T.~Petcov and A.~Y.~Smirnov,
Phys.\ Lett.\ B {\bf 322}, 109 (1994),
 arXiv:hep-ph/9311204;
 M.~Doi, T.~Kotani, H.~Nishiura, K.~Okuda and E.~Takasugi,
Phys.\ Lett.\ B {\bf 102}, 323 (1981).


\bibitem{MAP}
D.~J.~Eisenstein, W.~Hu and M.~Tegmark,
arXiv:astro-ph/9807130.




\end{thebibliography}
\end{document}